\documentclass[twocolumn]{aastex631}
\usepackage[capitalise]{cleveref}
\usepackage{hyperref}

\newcommand{\iap}{CNRS \& Sorbonne Universit\'{e}, Institut d’Astrophysique de Paris (IAP), UMR 7095, 98 bis bd Arago, F-75014 Paris, France}
\newcommand{\cca}{Center for Computational Astrophysics, Flatiron Institute, 162 5th Avenue, New York, NY 10010, USA}

\newcommand{\quijote}{\textsc{Quijote}}
\newcommand{\pylians}{\textsc{pylians}}

\submitjournal{ApJL}

\shorttitle{Bye bye, LIMD bias}
\shortauthors{Bartlett, Ho \& Wandelt}

\begin{document}

\title{Bye-bye, Local-in-matter-density Bias: The Statistics of the Halo Field Are Poorly Determined by the Local Mass Density}

\correspondingauthor{Deaglan J. Bartlett}
\email{deaglan.bartlett@iap.fr}

\author[0000-0001-9426-7723]{Deaglan J. Bartlett}
\affiliation{\iap}

\author[0000-0003-3207-8868]{Matthew Ho}
\affiliation{\iap}

\author[0000-0002-5854-8269]{Benjamin D. Wandelt}
\affiliation{\iap}
\affiliation{\cca}

\begin{abstract}

Bias models relating the dark matter field to the spatial distribution of halos are widely used in current cosmological analyses. 
Many models predict halos purely from the local Eulerian matter density, yet bias models in perturbation theory require other local properties.
We assess the validity of assuming
that only the local dark matter density can be used to predict the number density of halos in a model-independent way and in the non-perturbative regime.
Utilising $N$-body simulations, we study the properties of the halo counts field after spatial voxels with near-equal dark matter density have been permuted. If local-in-matter-density biasing were valid, the statistical properties of the permuted and un-permuted fields would be indistinguishable since both represent equally fair draws of the stochastic biasing model.
If the Lagrangian radius is greater than approximately half the voxel size
and for halos less massive than $\sim10^{15}\,h^{-1}{\rm\,M_\sun}$, we find the permuted halo field has a scale-dependent bias with greater than 25\% more power on scales relevant for current surveys.
These bias models remove small-scale power by not modelling correlations between neighbouring voxels, which substantially boosts large-scale power to conserve the field's total variance. This conclusion is robust to the choice of initial conditions and cosmology.
Assuming local-in-matter-density halo biasing cannot, therefore, reproduce the distribution of halos across a large range of scales and halo masses, no matter how complex the model. 
One must either allow the biasing to be a function of other quantities and/or remove the assumption that neighbouring voxels are statistically independent.

\end{abstract}



\section{Introduction}
\label{sec:Introduction}

Uncovering the statistical relation between the luminous tracers of large scale structure and the underlying matter density field -- the ``bias'' -- is a crucial part of cosmological analyses, since we observe the former but wish to understand the properties of the latter.
It has long been known that clusters form at high density peaks of the density field \citep{Kaiser_1984}, but an exact relation between these two quantities at the field level remains unknown.

On the largest scales, one expects that the bias is linear, namely the overdensity of tracers is proportional to the overdensity of matter. This approximation breaks down as one moves from linear to quasilinear scales, and more general functional relationships must be used. 
The simplest assumption one can make is that the tracer number density depends solely on the local Eulerian dark matter density - the so-called ``local-in-matter-density'' (LIMD) assumption \citep{Desjacques_2018} which, in a perturbative analysis, allows one to write the tracer number density as an expansion in the overdensity: the local bias expansion \citep{Fry_1993}.
More generally, one can express the bias model as local in gravitational observables, namely not just the density field but other terms related to the tidal field or higher derivative terms \citep{Mirbabayi_2015,Desjacques_2018,Lazeyras_2018,Lazeyras_2019}.
One can enumerate such terms based on the order of perturbation theory such that, if one goes to sufficiently high order and is in the large-scale, perturbative regime, then the result converges to the truth.

Relying on perturbation theory \citep{Bernardeau_2002} or effective field theory (EFT) \citep{Carrasco_2012,Senatore_2014,Senatore_2015,Perko_2016} inherently limits our analysis to these quasi-linear scales.
Although it is possible to study higher order statistics \citep{D_Amico_2022} or perform field-level analyses in these frameworks \citep{Schmidt_2019,Elsner_2020,Kostic_2023,Stadler_2023,Nguyen_2024}, to extract information on non-linear scales one may need to go beyond such local bias expansions, although some studies have shown that perturbative methods perform well when tested against $N$-body simulations \citep{Roth_2011,Schmittfull_2019}.
Instead of relying of power series, many parametric forms for the biasing relation have been proposed in the literature based on the LIMD assumption \citep{Szalay_1998,Matsubara_1995,Matsubara_2011,Frusciante_2012,Neyrinck_2014,Ata_2015}.
As such, many reconstruction analyses typically assume that the number density of halos is solely a function of the smoothed density field 
\citep[e.g.][]{Schmoldt_1999,Erdogdu_2004,Erdogdu_2006,Kitaura_2009,Jasche_2010,Jasche_2012,Jasche_2013,Jasche_2015,Lavaux_2016,Modi_2018,Jasche_2019,Lavaux_2019,Ramanah_2019}.
Alternatively, one can utilise machine learning methods to map the dark matter density field (using both local and non-local information) to the halo counts field (\cite{Charnock_2020}; see \cite{Dai_2021} for a Lagrangian approach).

Unlike in the perturbative regime, we do not have theoretical guarantees that any of these bias models are accurate and so this must be tested empirically. 
\citet{Mirbabayi_2015} argue that, at any order in perturbation theory, the bias can be expressed in terms of locally measurable observables, and that this argument is closed under renormalisation. However, it could be (as is often assumed \citep{Szalay_1998,Matsubara_1995,Matsubara_2011,Frusciante_2012,Neyrinck_2014,Ata_2015}) that the only important observable is the local dark matter density field. Testing this assumption is the subject of this letter.
There is an important distinction, however, between testing the assumptions of the halo biasing model (i.e. which variables are needed to predict the tracer number density) and the particular model chosen (i.e. what is the functional form of the bias or the stochasticity in the tracer-matter relation).
Previous tests of the latter seem to suggest that information beyond the local density is required 
\citep[e.g.][]{Nguyen_2021,Fang_2024}, 
however this could just be due to insufficiently accurate LIMD biasing models. Initial studies of the former also seem to support this conclusion \citep{Balaguera-Antolinez_2019,Pellejero-Ibanez_2020,Balaguera-Antolinez_2020}.

In this letter we study the former issue in the fully non-linear regime.

We present a model-independent test of the assumption of LIMD halo biasing. Our test is independent of the functional form of the biasing model, but simply tests whether halo biasing can be a stochastic function of only the local matter density.
We find that LIMD halo biasing cannot reproduce the distribution of halos less massive than $\sim 10^{15} \, h^{-1} \, {\rm M_\sun}$ when applied to gridded dark matter fields, when the Lagrangian radius (\cref{eq:Lagrangian radius}) of the halo is larger than approximately half the voxel size.
Our permutation test naturally enforces conservation of all 1-point moments and count statistics, while demonstrating the insufficiency of the LIMD hypothesis to model halo clustering.
We demonstrate that this conclusion is robust across initial conditions and cosmology by utilising the \textsc{quijote} suite of $N$-body simulations.
We conclude that no LIMD biasing model, however complex, should be used if one wishes to accurately predict the spatial distribution of halos given a matter density field.

Our test also provides a method for testing the convergence of halo catalogues from $N$-body simulations; we find that the minimum mass for which the catalogues are converged for \quijote{} is $\sim10^{13.1} \, h^{-1} {\rm \, M_\sun}$.

In Sec.~\ref{sec:Methods} we explicitly define LIMD biasing, introduce our model-independent test and describe the simulations used to perform this test. Our results are presented in Sec.~\ref{sec:Results}, where we demonstrate that our conclusions are highly robust to the one free parameter in our test, as well as cosmology and initial conditions, and investigate the adequacy of LIMD biasing as a function of scale and halo mass. Sec.~\ref{sec:Conclusion} discusses these results and presents our conclusions.

\section{Methods} 
\label{sec:Methods}

\subsection{Assessing the assumption of local biasing}

We define local-in-matter-density (LIMD) halo biasing to be the assumption that the number of halos in a given region of comoving space (voxel) is drawn from any probability distribution under the assumptions:
\begin{enumerate}
    \item The parameters of this distribution in a given voxel are only functions of the dark matter density in that voxel.
    \item Each voxel is statistically independent.
\end{enumerate}
Note that we will not make any assumptions about the type of distribution (e.g. Poisson) nor the functional form of its parameters (e.g. the mean for a Poisson distribution) other than we expect that this function should be a continuous function of the local dark matter density.
The statement about statistical independence is made within the context of conditional probability. Specifically, while the underlying quantities that influence the distribution in each voxel are indeed correlated among voxels (inherited from the correlation due to large-scale structure), the stochastic quantities drawn from this distribution at each voxel are treated as independent random variables.
If the particular draw from the conditional distribution depended on other voxels in any other way than through this inherent large-scale structure correlation, then one would need information beyond the local density field to reconstruct the halo counts field, and thus this would not be a model which is not local in matter density.

Under these assumptions, if we identify several voxels which have the same dark matter density and then permute the number of halos in these voxels, we should end up with an equally likely draw from the underlying distribution. If the statistical properties of the halo-count field are noticeably different after this permutation, then at least one of the assumptions of LIMD biasing cannot be true.

Since the dark matter density field takes continuous values, instead of performing permutations on voxels which have exactly equal dark matter density, we first sort the voxels into $n_{\rm bin}$ bins of dark matter density, and consider two voxels equivalent if they fall within the same bin, namely if their density is approximately equal.
In our fiducial analysis, we choose to use adaptive bin widths so that each bin has approximately the same number of voxels and use $n_{\rm bin}=1000$.
We demonstrate that our results are highly robust to this choice in Sec.~\ref{sec:Results}.

To assess the level of agreement between the permuted and un-permuted fields, we compute the power spectrum, $P_{\rm counts}(k)$, of the counts field using the \pylians{} library \citep{Pylians}. We perform 100 permutations and report the mean and standard deviation at each wavenumber, then compare this to the true $P_{\rm counts}(k)$.

\subsection{Simulations}

To test the assumption of LIMD biasing, we utilise the \quijote{} suite of $N$-body simulations. These simulations were run in a periodic box of length $L = 1 \, h^{-1} {\rm Gpc}$ using the \textsc{gadget-iii} code \citep{Springel_2005}. We focus on the high-resolution suite which contain $N_{\rm p}^3=1024^3$ particles, as opposed to the standard resolution of $N_{\rm p}^3=512^3$.

For our fiducial analysis, we consider simulation number 0 from the fiducial cosmology, which matches the Planck 2018 cosmological parameters \citep{Planck_VI_2018}: $\Omega_{\rm m} = 0.3175$, $\Omega_{\rm b} = 0.049$, $h=0.6711$, $n_{\rm s} = 0.9624$, $\sigma_8=0.834$.
In Sec.~\ref{sec:Results} we verify that our conclusions are robust if we chose different initial conditions (i.e. for different simulation numbers in the suite). We also study the sensitivity of our conclusions to the choice of cosmological parameters by performing our analysis with the (high-resolution) Latin-hypercube suite of \quijote{} simulations, consisting of 2000 dark-matter-only simulations with cosmological parameters arranged on a Latin hypercube in the range: $\Omega_{\rm m} \in [0.1, 0.5]$, $\Omega_{\rm b} \in [0.03, 0.07]$, $h\in [0.5, 0.9]$, $n_{\rm s} \in [0.8, 1.2]$, $\sigma_8 \in [0.6, 1.0]$.

We make all our comparisons at redshift zero, where we use the positions and masses of halos from the pre-computed Friends-of-Friends (FoF) \citep{Davis_1985} halo catalogues, which were obtained using a linking length of $b=0.2$.
For brevity, throughout this letter we only report results using FoF halos, but we have verified that our conclusions are robust to this choice by rerunning our analysis with the Rockstar halofinder \citep{Behroozi_2013}, and found consistent conclusions. 

Since one would expect halos of different masses to be distributed differently in space, we binned these halos into nine logarithmically spaced mass bins in the range $10^{13.1}-10^{15.8} \, h^{-1} \, {\rm M}_\odot$, and performed our test separately for each bin. 
For halos within this mass range, we find that our results are consistent between the standard and high-resolution \quijote{} simulations, whereas this is not true for lower masses. We therefore cannot be confident that the halo catalogues are converged for halos which are less massive than this, so simply discard them in our analysis.
Our test, therefore, additionally acts as a method of determining the minimum mass for which the halo catalogues from these simulations are converged.

We divide the simulation volume into $N^3$ cubic voxels and compute the density field in each voxel 
using a nearest grid point (NGP) estimator. This is the same estimator used to compute the gridded halo counts field, and thus we mitigate effects which arise due to different discretisation procedures. We have tested our method using a range of other density field estimators: cloud in cell (CIC), triangular-shape cloud (TSC), and piecewise cubic spline (PCS) kernel, as well as one inspired by smoothed particle hydrodynamics (SPH) \citep{Monaghan_1992,Colombi_2007}.
For our fiducial analysis we used $N=128$, corresponding to voxels with a side length of $\Delta x = 7.8 h^{-1} {\rm Mpc}$, but we show that our results are insensitive to this choice (and thus to the smoothing scale of the density field) in Sec.~\ref{sec:Results}.
At our fiducial grid resolution, we find that all density estimators give the same qualitative trends, although the magnitude of the bias induced by LIMD models varies with this choice. 
For large voxel sizes, we find the performance is best for the NGP density estimator, but the degree of bias depends sensitively on the choice of mass assignment in the definition of the density field and is much worse for other choices.
We thus choose the NGP estimator to ensure our results are conservative.

\section{Results} \label{sec:Results}

\subsection{LIMD halo biasing cannot reproduce the halo distribution}

\begin{figure}
    \centering
    \includegraphics[width=\columnwidth]{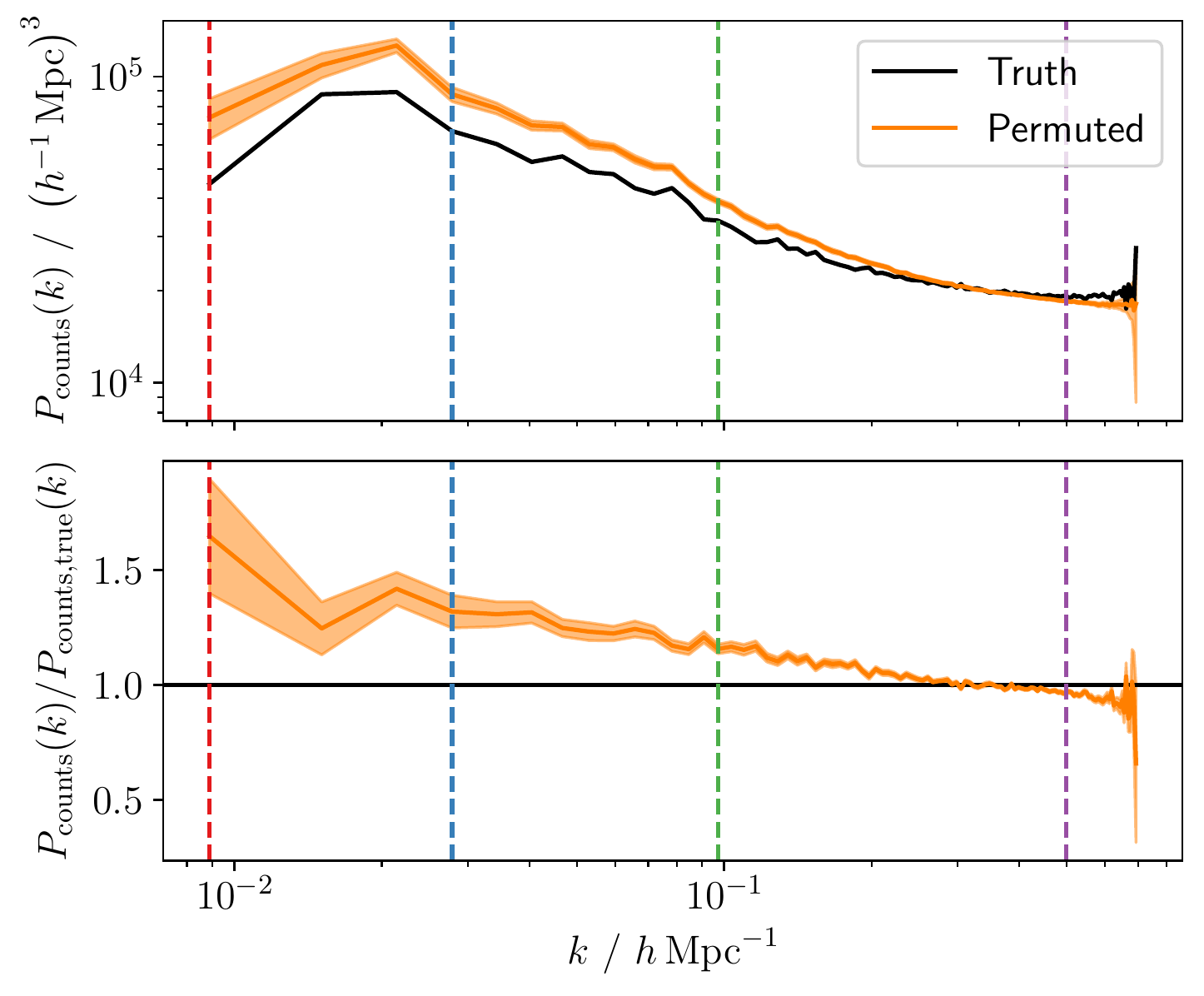}
    \caption{Power spectrum of the halo counts field before and after permuting voxels with approximately equal density, for halos of mass $10^{13.7}-10^{14.0} \, h^{-1} {\rm \, M}_\sun$ . The permuted fields have significantly greater power on large scales, thus LIMD halo biasing cannot reproduce the distribution of halos. The mean and standard deviation are computed over 100 permutations. The vertical coloured lines correspond to the wavenumbers used to characterise this discrepancy in later plots.}
    \label{fig:pk_comparison}
\end{figure}

In \cref{fig:pk_comparison} we plot $P_{\rm counts}(k)$ for our fiducial setup for halos in the mass range $10^{13.7}-10^{14.0} \, h^{-1} {\rm \, M}_\sun$, both before and after permuting voxels with equal dark matter density. 
When compared to the truth, we see that the permuted fields have very different halo distributions than the original field on all scales. In particular, we observe a significant excess of power on scales $k \lesssim 0.3 \, h \, {\rm Mpc^{-1}}$ and too little power on smaller scales.
We infer that this is due to removal of small-scale power, due to LIMD models ignoring local correlations, which then requires an overestimation of large-scale power, to respect the conservation of the field-variance under voxel permutations.
This represents the main conclusion of this letter: the assumption of LIMD halo biasing cannot produce the correct distribution of halos.

The vertical dashed lines in \cref{fig:pk_comparison} correspond to four characteristic wavenumbers which we use for the remainder of the analysis to quantify the level of discrepancy, corresponding to approximately $0.01$, $0.03$, $0.1$ and $0.5 \, h \, {\rm Mpc^{-1}}$. For our fiducial analysis, we find that $P_{\rm counts}(k)$ is a factor of 
$1.6\pm0.2$, $1.3\pm0.07$, $1.16\pm0.02$ and $0.97\pm0.01$
times the true $P_{\rm counts}(k)$, respectively, for these wavenumbers. It is therefore clear that the discrepancy is significant for this simulation, with the true power spectrum at approximately the $7.6\sigma$ value for $k=0.1 \, h \, {\rm Mpc^{-1}}$.

\begin{figure}
    \centering
    \includegraphics[width=\columnwidth]{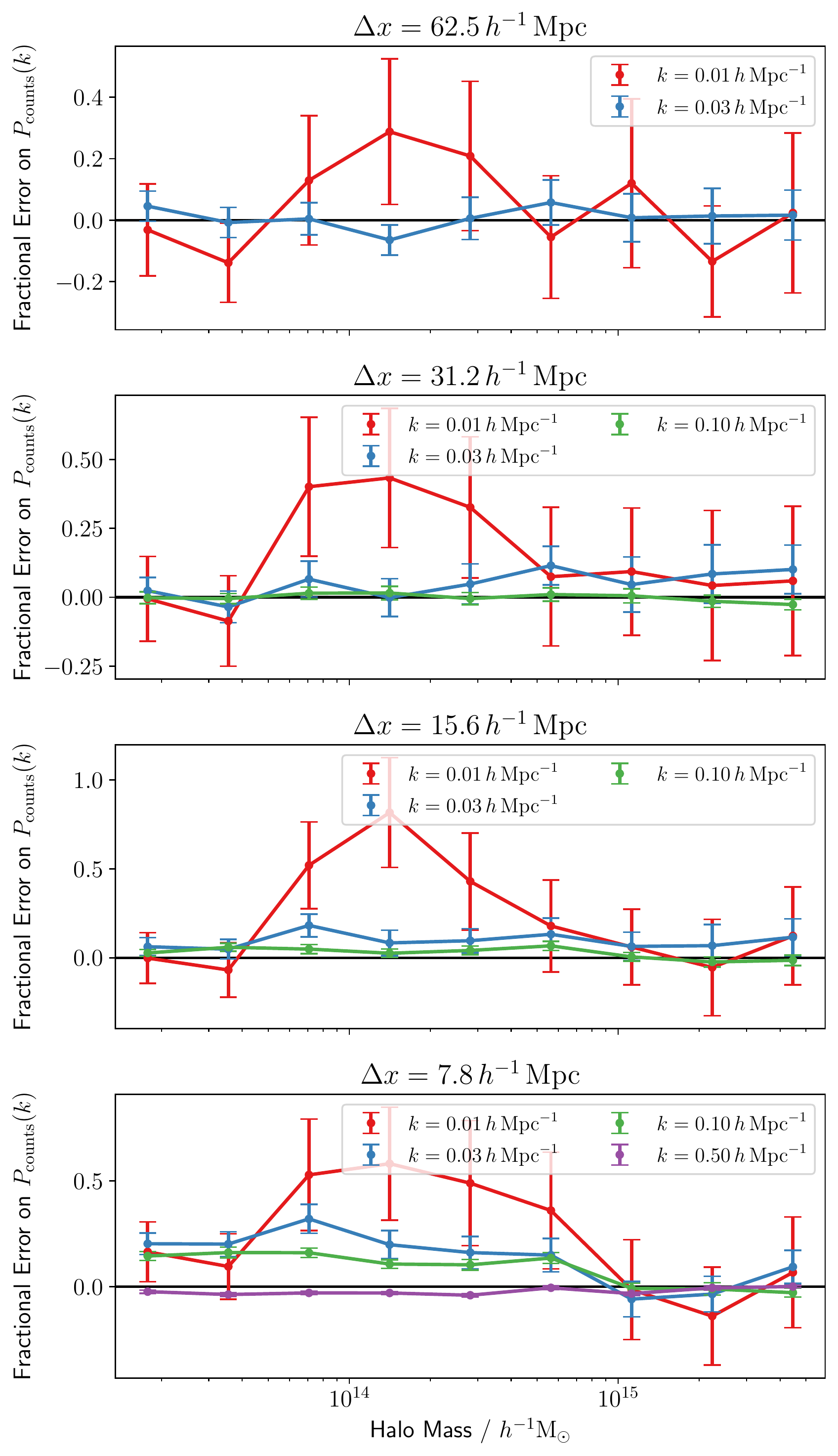}
    \caption{Fractional error on the power spectrum after permuting equal-density voxels as a function of halo mass and voxel size, $\Delta x$. Significant discrepancies are seen for a wide range of scales for halo masses below $10^{15} \, h^{-1} {\rm \, M}_\sun$ and when $\Delta x \lesssim 30 \, h^{-1}{\rm \, Mpc}$, hence LIMD halo biasing is invalid in this regime. The points and error bars give the mean and standard deviation of the discrepancy across 100 permutations, respectively.}
    \label{fig:mass_comparison}
\end{figure}

In \cref{fig:mass_comparison} we investigate how this conclusion changes as a function of halo mass and voxel size. 
For the smaller voxel sizes ($\Delta x \lesssim 30 \, h^{-1}{\rm \, Mpc}$), one observes that the fractional difference is non-zero until
$\sim 10^{15} \, h^{-1} {\rm \, M}_\sun$, indicating that LIMD halo biasing is invalid for all masses below this point. For the largest halo masses and voxels, we find that the power spectrum of the permuted field is consistent with the truth. The number density of these objects is low -- there are only 381 halos above $10^{15} \, h^{-1} {\rm \, M}_\sun$ in the fiducial simulation -- and these can only exist at the extreme peaks of the density field. Hence, given that these objects cannot exists in low density environments, it is perhaps unsurprising that their number density can be well-predicted by just the local density field. We therefore conclude that LIMD halo biasing is a reasonable assumption for halos above $\sim 10^{15} \, h^{-1} {\rm \, M}_\sun$, but, for the majority of objects in the simulation, this framework cannot be used when the voxel size is below $\sim 30 \, h^{-1}{\rm \, Mpc}$.

\begin{figure}
    \centering
    \includegraphics[width=\columnwidth]{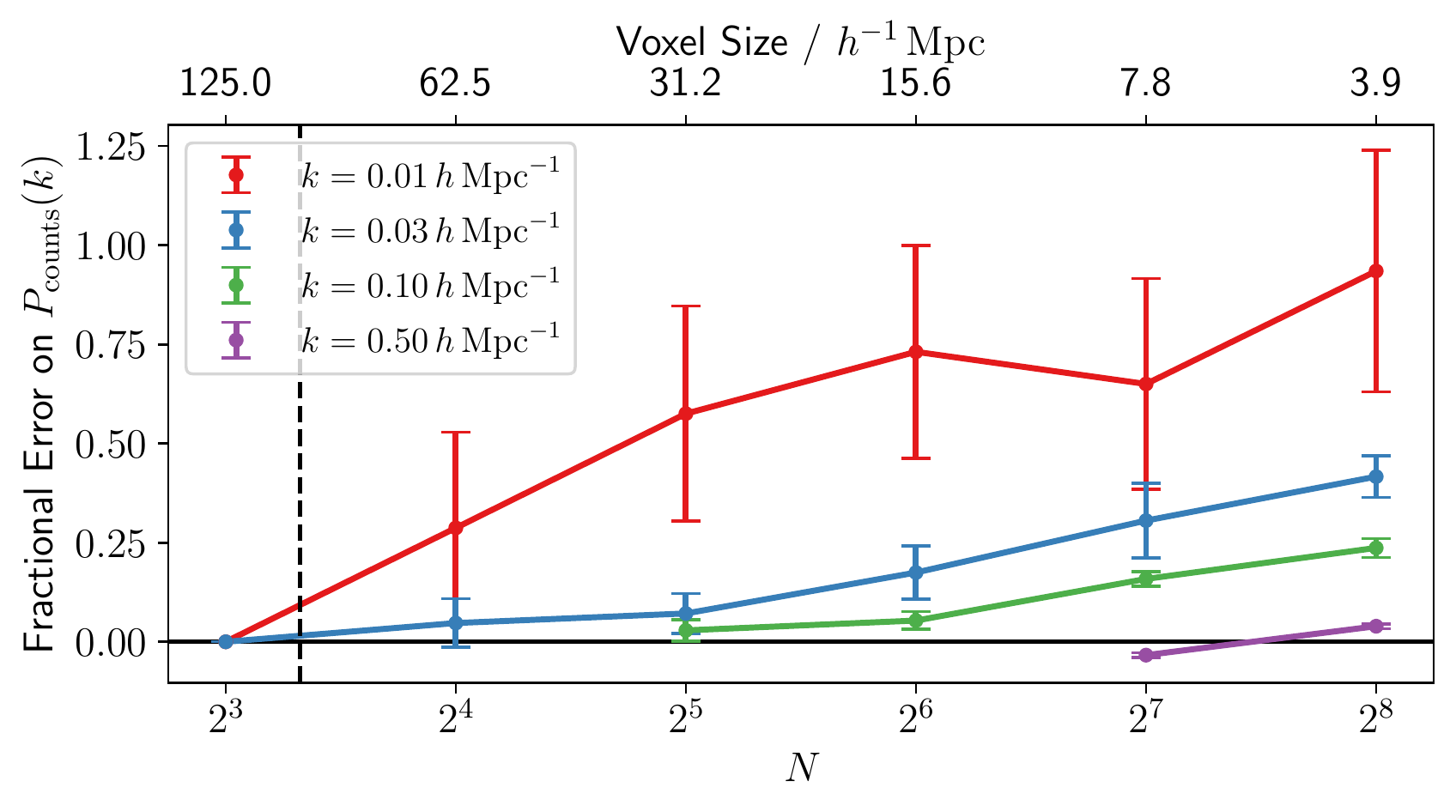}
    \caption{Fractional error on the power spectrum for halos of mass $10^{13.7}-10^{14.0} \, h^{-1} {\rm \, M}_\sun$ after permuting equal-density voxels as a function of voxel size. The black dashed line indicates the value of $N$ where we have as many voxels as density bins. For $N\geq 2^6$ we find significant disagreement on large scales, suggesting that one must smooth the density field over at least $30 \, h^{-1} {\rm \, Mpc}$ for LIMD halo biasing to be valid. We do not plot values for wavenumbers beyond the Nyquist frequency, hence some lines are truncated.}
    \label{fig:N_comparison}
\end{figure}
 
In \cref{fig:N_comparison} we investigate how the error on $P_{\rm counts}(k)$ varies as a function of grid size for our fiducal mass bin. 
One observes that the level of disagreement is relatively unchanged on the largest scales as one changes the voxel size from $\sim 4 \, h^{-1}{\rm \, Mpc}$ to $\sim 30 h^{-1}{\rm \, Mpc}$, 
indicating that LIMD biasing is inappropriate for density fields smoothed on these scales. As one further increases the voxel size to $\sim 60 h^{-1}{\rm \, Mpc}$, the bias starts to diminish, with a bias of $\sim1.4\sigma$ at $k = 0.03 \, h \, {\rm Mpc^{-1}}$. 
In this regime the number of voxels is similar to the number of density bins (as indicated by the black line in \cref{fig:N_comparison}), but we have verified that similar behaviour is seen if we reduce $n_{\rm bin}$ to 100.
However, even on scales $\sim 0.01 \, h{\rm \, Mpc}^{-1}$, we see a mean fractional error of 25\% for a voxel size of $\sim 62.5 \, h^{-1}{\rm \, Mpc}$.
Thus, the LIMD halo biasing is invalid for smoothed density fields with a smoothing scale smaller than $\sim 30 h^{-1}{\rm \, Mpc}$, although it may be an acceptable approximation on larger scales.

\begin{figure}
    \centering
    \includegraphics[width=\columnwidth]{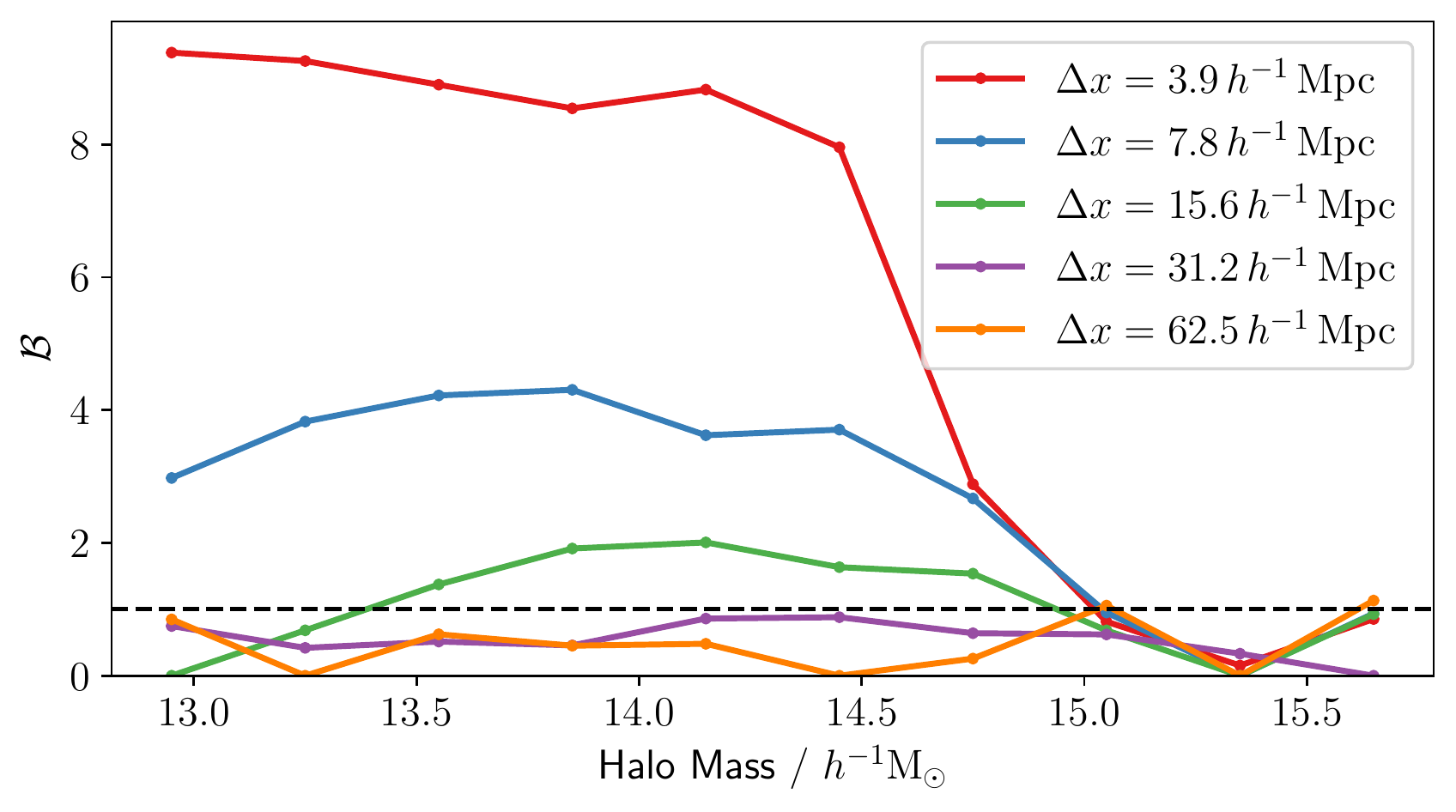}
    \caption{The significance of the deviation (\cref{eq:summary}) between the power spectra of the true and permuted fields as a function of halo mass, for different voxel sizes, $\Delta x$. The dashed line corresponds to $\mathcal{B}=1$. Large values of $\mathcal{B}$ correspond to scenarios where LIMD is a poor approximation.}
    \label{fig:summary_comparison}
\end{figure}

To further investigate how the error on $P_{\rm counts}(k)$ varies as a function of grid size and mass, we first define a way of summarising the bias across the different wavenumbers. If the fractional error at each $k$ is $\epsilon(k)$ with standard deviation $\sigma(k)$, then we define
\begin{equation}
    \label{eq:summary}
    \mathcal{B}^2 =  \frac{\int \left( \frac{\epsilon(k)}{\sigma(k)} \right)^2 {\rm d} k}{\int {\rm d} k} - 1.
\end{equation}
If the fractional error had a constant mean of $n \sigma(k)$ for all wavenumbers and each wavenumber were independent, then this gives $\langle \mathcal{B}^2 \rangle = n^2$, and thus $\mathcal{B}$ gives a dimensionless estimate of the bias, where $\mathcal{B}=0$ corresponds to being unbiased.

We compute this quantity for the various mass bins and resolutions, and plot the resulting $\mathcal{B}$ as a function of halo mass in \cref{fig:summary_comparison}. As before, we find that the level of bias is small for $\Delta x \gtrsim 30 h^{-1}{\rm \, Mpc}$, with $\mathcal{B} < 1$ for all halo masses.
For the most massive halos (above $\sim 10^{15} \, h^{-1} {\rm \, M}_\sun$), we again see no indication of bias for any grid resolution, presumably since these objects are rare and can only exist in the densest environments.
As we move to smaller masses and $\Delta x < 30 \, h^{-1}{\rm \, Mpc}$, we find that the bias increases, with $\mathcal{B} > 8$ for $\Delta x = 3.9 \, h^{-1}{\rm \, Mpc}$. 

Intriguingly, we find that the value of $\mathcal{B}$ does not monotonically increase as one decreases the halo mass, but for $\Delta x = 15.6 \, h^{-1}{\rm \, Mpc}$, $\mathcal{B}$ reaches a maximum of $\sim 2$ at $\sim 10^{14.2} \, h^{-1} {\rm \, M}_\sun$, before decreasing to below one at masses below $\sim 10^{13.4} \, h^{-1} {\rm \, M}_\sun$. One can begin to see the same effect for $\Delta x = 7.8 \, h^{-1}{\rm \, Mpc}$, with the maximum at $\sim 10^{13.5} \, h^{-1} {\rm \, M}_\sun$. These maxima occur when the voxel size is approximately twice the Lagrangian radius of the halo, 
defined such that the halo mass, $M_{\rm h}$, obeys 
\begin{equation}
    \label{eq:Lagrangian radius}
    M_{\rm h} \equiv \frac{4 \pi}{3} \rho_{\rm m} R_{\rm L}^3,
\end{equation}
where $\rho_{\rm m}$ is the cosmic mean matter density. Therefore, one would expect to see a similar maximum for $\Delta x= 3.9 \, h^{-1}{\rm \, Mpc}$ for a halo mass of $\sim 10^{12.4} \, h^{-1} {\rm \, M}_\sun$.
However, these halos are not resolved in the \quijote{} simulations, so we cannot test this conjecture here. As one decreases the halo mass below this maximum, the voxel size becomes significantly larger than the Lagrangian radius of the halo, and thus LIMD biasing becomes an increasingly good approximation. If the Lagrangian patch is larger than the voxel size, then information from neighbouring voxels is required to determine the number and positions of halos.
Hence, we conclude that there exists an intermediate range of masses for a given voxel size where LIMD biasing breaks down, where the upper limit is due to the rarity of very massive halos, and the lower limit is set by the ability to resolve the Lagrangian radius of individual halos.

\subsection{Sensitivity to binning scheme}

\begin{figure}
    \centering
    \includegraphics[width=\columnwidth]{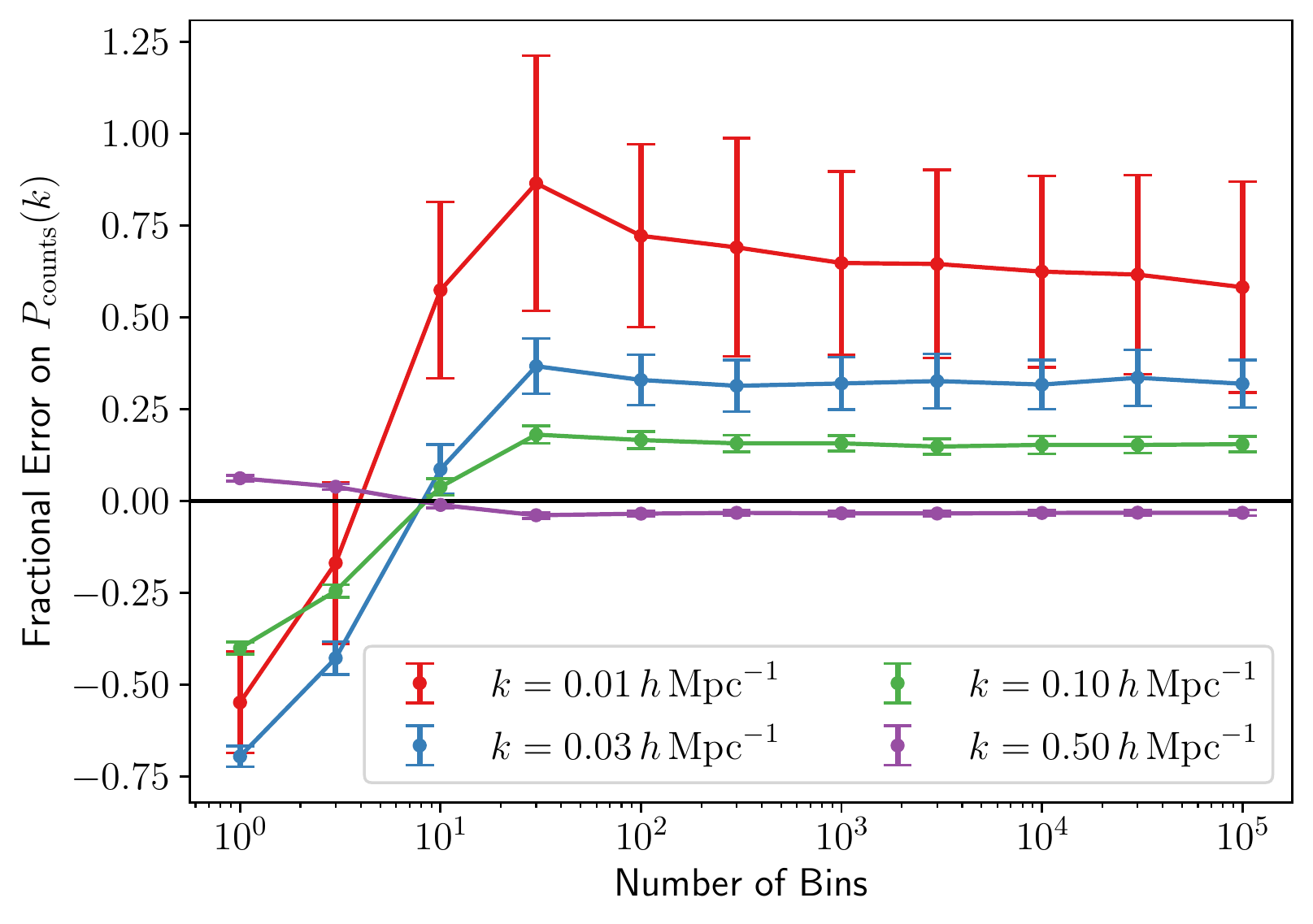}
    \caption{Fractional error on the power spectrum of halo counts after permuting voxels within the same density bin as a function of the number of bins. We consider halos of mass $10^{13.7}-10^{14.0} \, h^{-1} {\rm \, M}_\sun$ and choose a binning scheme such that each bin contains approximately an equal number of voxels. After an initial transient behaviour, we find constant biases in the power spectrum across many order of magnitude, hence our conclusion is robust to the choice of $n_{\rm bin}$.}
    \label{fig:nbin_comparison}
\end{figure}

If one did not bin the density sufficiently finely, then one would expect that this shuffling procedure would not reproduce the correct distribution of halos even if LIMD halo biasing were correct, since voxels with very different densities would be considered equivalent and therefore permuting their halos would not represent any realistic biasing model. 
To test this, in \cref{fig:nbin_comparison} we plot the fractional error on the power spectrum as a function of $n_{\rm bin}$.
When one has only one density bin, the permuted-halo power spectrum is consistent with random noise.
As we increase $n_{\rm bin}$, one sees an initial transient behaviour, where the fractional error becomes positive on large scales at approximately $n_{\rm bin}=10$ and then increases to a constant level by approximately $n_{\rm bin}=30$.
Beyond this point, the fractional error on $P_{\rm counts}(k)$ is independent across orders of magnitude of $n_{\rm bin}$ suggesting that our conclusions are robust to this choice.

We have verified that our results are unchanged if we consider equally spaced bins in both density and the logarithm of density, instead of enforcing each bin to have approximately an equal number of voxels. The transient behaviour disappears at $n_{\rm bin} \approx 10$ when using logarithmic bins and at $n_{\rm bin}\approx300$ for the linear binning scheme. Above these values, we again find that the bias is insensitive to the choice of $n_{\rm bin}$. 
We have also verified that our binning is sufficiently fine such that shuffling the density field under this scheme yields a matter power spectrum which is consistent with the unshuffled field.

Of course, if one had as many bins as number of voxels ($n_{\rm bin} = N^3)$, then (assuming no voxels have identically equal densities) there would be only one voxel per bin and thus the permuted halo field is equal to the unpermuted field. For $N=128$ this corresponds to $n_{\rm bin} \approx 10^{6.3}$.
Since our conclusion holds at $n_{\rm bin}=10^5$, even with an average of just 21 voxels per bin we cannot reproduce the halo distribution. 
A LIMD halo biasing model which varies that rapidly with density seems implausible and highly fine-tuned.
We have verified that using $n_{\rm bin}=100$ for a mock halo counts field generated from a Poisson distribution with a mean which depends on density according to a power law (with parameters optimised to fit the \quijote{} simulations using a Poisson loss function in each voxel) yields unbiased $P_{\rm counts}(k)$ upon shuffling, again indicating that our fiducial choice of $n_{\rm bin}=10^3$ is sufficient.

\subsection{Sensitivity to cosmology and initial conditions}

To test the robustness of our conclusion to the initial conditions, we repeated our analysis across the 100 high-resolution \quijote{} simulations at the fiducial cosmology. We find biases (mean divided by standard deviation across 100 permutations) of 
$1.4 \pm 0.9$, $4.5 \pm 1.0$, $6.6 \pm 0.9$, $-4.7 \pm 1.0$
at the wavenumbers shown in \cref{fig:pk_comparison}, indicating that our conclusion is insensitive to the initial white noise field.

Similarly, to investigate the sensitivity of the result to the cosmological parameters, we ran our analysis for all 2000 high-resolution simulations in the Latin hypercube suite. We find that, averaging over cosmological parameters, the biases for these wavenumbers are $1.7 \pm 1.4$, $4.2 \pm 1.5$, $6.4 \pm 1.2$, $-4.5 \pm 1.1$
demonstrating that LIMD halo biasing does not hold for any of the cosmologies considered.

\section{Discussion and conclusion}
\label{sec:Conclusion}

In this letter we have shown that the distribution of halos cannot be modelled according to LIMD biasing: the assumption that the number of halos in a given voxel is drawn from a probability distribution which depends solely on the local dark matter density and where each voxel is statistically independent. At least one of these assumptions must be broken to accurately model the distribution of halos in the Universe when averaging over scales of $\sim 4-30 \, h^{-1} {\rm \, Mpc}$.
Of course, one would expect that if one went to sufficiently small scales, halos would extend to multiple voxels and thus LIMD biasing must break once the voxels are made sufficiently small. This work quantifies this scale and its relation to the halo's Lagrangian radius (\cref{eq:Lagrangian radius}) using $N$-body simulations.

We have demonstrated this in a model-independent way, without considering a particular choice of probability distribution nor the dependence of its parameters on the local dark matter density. By permuting the number of halos found in voxels of approximately equal density, we showed that the permuted field has too much power on large scales, whereas this would be statistically indistinguishable from the un-permuted field if LIMD biasing were true. This conclusion is robust to the definition of ``approximately equal density'' as well as voxel size, cosmology and initial conditions. We find that the most massive halos can be modelled under these assumptions, but this breaks down for halos less massive than $\sim 10^{15} \, h^{-1} {\rm \, M_\sun}$.

The test described in this letter is similar in approach to the first step of the Bias Assignment Method (BAM) \citep{Balaguera-Antolinez_2019}, which is used to fit non-parametric bias models. As here, \citet{Balaguera-Antolinez_2019} find that information beyond the local density is required to recover the halo distribution but perform their analysis with a single grid resolution of $\Delta x = 3 \, h^{-1} {\rm \, Mpc}$, which is slightly finer than we consider. Our work extends this conclusion by systematically exploring the sensitivity to this smoothing scale and explicitly investigating the mass dependence of this conclusion in separate mass bins. BAM has recently been used to provide evidence for non-locality in the bias model by studying the bispectrum \citep{Pellejero-Ibanez_2020}, with up to $4.8\sigma$ significance \citep{Kitaura_2022}. The conclusions presented here are much stronger than this. Moreover, we demonstrate that the results are robust to cosmological parameters and provide more thorough tests of the robustness of our results to the binning scheme. Unlike \citet{Balaguera-Antolinez_2019}, we have demonstrated that our conclusions are robust to the resolution of the $N$-body simulation used in our test.

This conclusion is particularly important as the use of higher-order statistics and field-level approaches become increasingly popular in cosmological analyses, since it is no longer sufficient to tune a bias model to only reproduce the power spectrum or other low-order statistics; one must be able to reproduce all moments of the field.
In field-level studies, the initial conditions of the Universe are free parameters, so using a LIMD biasing model implies inferring an incorrect density field to obtain a predicted galaxy (or whichever tracer is used in the analysis) field which reasonably matches observations \citep{Nguyen_2021}. 
Although the inferred galaxy field at the redshift of the observations would be trustworthy, extrapolating to other redshifts could be problematic and the underlying density field and its initial conditions would have artefacts.
Given that it is often these initial conditions which one is interested in, it is imperative that future studies move beyond the LIMD halo biasing assumption to avoid these problems.

If the issue causing the deficiency of LIMD models was chiefly the dependence on local density (i.e. independent draws is a reasonable approximation), then one could apply a similar permutation technique to identify whether other halo biasing schemes are valid in a model-independent fashion. For example, by binning in multiple dimensions, one could assess whether the variables used in EFT based approaches are sufficient in the non-perturbative regime.
Alternatively, one could search for transformations of the density field for which the permuted and un-permuted halo number counts are indistinguishable when binned in this variable and then construct bias models with these new parameters.

As discussed in Sec.~\ref{sec:Methods}, the degree to which LIMD biasing fails depends on the estimator used to obtain the density field. This high sensitivity to the estimator is problematic if one wishes to predict the halo distribution from an approximate gravity model. As such, using information beyond the local density field is essential to produce realistic halo catalogues for analysis pipelines for current and future surveys.

We conclude that number of halos in a given region of space is a stochastic function of the local dark matter density field, but it cannot depend solely on this quantity.
Models which rely on this assumption, no matter how complex, cannot reproduce the distribution of halos and thus must be superseded by more accurate models.
This letter not only disproves the LIMD assumption of halo biasing for scales 
which are not significantly greater than the Lagrangian radius of a halo,
but provides a framework to aid the design of future approaches by testing their assumptions in a model-independent manner.

\begin{acknowledgments}
We thank Andrés Balaguera, Yan-Chuan Cai, Neal Dalal, Simon Ding, Ludvig Doeser, Rozo Eduardo, Pedro Ferreira, Joachim Harnois-Deraps, Jens Jasche, Guilhem Lavaux, Titouan Lazeyras, Stuart McAlpine, Minh Nguyen, Cora Uhlemann, Francisco Villaescusa-Navarro and the Implicit Likelihood Inference, Accelerated Forward Modelling and Robustness working groups of the Simons Collaboration on ``Learning the Universe'' for useful comments and suggestions.

DJB and MH are supported by the Simons Collaboration on ``Learning the Universe.''
BDW acknowleges the DIM-ORIGINES-2023, \textit{Infinity Next} grant. The Flatiron Institute is supported by the Simons Foundation.
This work was done within the Aquila Consortium (\url{https://www.aquila-consortium.org/}).

For the purposes of open access, the authors have applied a Creative Commons Attribution (CC BY) licence to any Author Accepted Manuscript version arising.
\end{acknowledgments}

\software{
\textsc{astropy} \citep{astropy_1,astropy_2,astropy_3},
\textsc{matplotlib} \citep{matplotlib},
\textsc{numpy} \citep{numpy},
\textsc{pandas} \citep{pandas_paper,pandas_zeondo},
\pylians{} \citep{Pylians}.
          }

\bibliography{references}{}
\bibliographystyle{aasjournal}


\end{document}